\documentclass[a4paper,twoside,12pt]{article}
\usepackage{epsfig}
\def \pom {{\hspace{ -0.1em}I\hspace{-0.2em}P}}
\newcommand{\lsim}{\raisebox{-0.5mm}
{$\stackrel{<}{\scriptstyle{\sim}}$}}
\newcommand{\ffigh}[4]{\begin{figure}[htbp]\vfill\begin{center}
\mbox{\epsfig{figure=#1,height=#2}}\caption{#3}\label{#4}
\end{center}\vfill\end{figure}}
\newcommand{\ffigw}[4]{\begin{figure}[htbp]\vfill\begin{center}
\mbox{\epsfig{figure=#1,width=#2}}\caption{#3}\label{#4}
\end{center}\vfill\end{figure}}

\newcommand{\feynfig}[4]{\begin{figure}[htbp]\vfill\begin{center}
\mbox{\epsfig{figure=#1,height=#2}}\vspace{0.3cm}\caption{#3}\label{#4}
\end{center}\vfill\end{figure}}

\begin{document}
\noindent
DESY 96-060 \hfill ISSN 0418-9833\\
April 1996\\
\begin{center}
 \begin{Large}
 \begin{bf}
   Unified Description of Rapidity Gaps\\
   and Energy Flows in DIS Final States \\
 \end{bf}
 \end{Large}
 \vspace*{5mm}
 \begin{large}
A.~Edin$^1$, G.~Ingelman$^{1,2}$, J.~Rathsman$^1$\\
 \end{large}
\vspace*{3mm}
 {\it $^{1}$Dept. of Radiation Sciences, Uppsala University, 
 Box 535, S-751 21 Uppsala, Sweden}\\
 {\it \mbox{$^{2}$Deutsches Elektronen-Synchrotron DESY, 
 Notkestrasse 85, D-22603 Hamburg, Germany}}\\
\vspace*{5mm}
\end{center}
\begin{quotation}
\noindent
{\bf Abstract:}
We show that the `orthogonal' characteristics of the observed rapidity gaps 
and large forward energy flows in deep inelastic scattering at HERA, can be 
described within a single framework. 
Our Monte Carlo model is based on perturbative QCD matrix elements 
and parton showers together with Lund string model hadronization, but has 
 in addition a new mechanism for soft colour interactions which 
modifies the perturbative colour structure and thereby the hadronization. 
Effects of perturbative multiparton emission are investigated and the 
non-perturbative treatment of the proton remnant is discussed and comparison 
to the observed transverse energy flow is made. We investigate
the resulting diffractive-like properties of the model; such as rapidity 
gap events, $t$- and $M_X$-distributions and the diffractive structure 
function in comparison to H1 data.
\end{quotation}

\section{Introduction}
\label{sec:intro}

The hadronic final state in deep inelastic scattering (DIS) of leptons on 
nucleons provides information on strong interaction dynamics. The availability
of many different observables, such as particle and energy flows, jet 
structures etc., give possibilities for investigations of perturbative and 
non-perturbative QCD in extension to those based on measurements of 
inclusive structure functions.  

Valuable results from final state investigations have been obtained at various 
fixed target experiments and with the electron-proton collider HERA at DESY
a large new kinematic domain has been made available. In particular, the much
larger range in mass $W$ of the hadronic system, extending up to the kinematic 
limit $\sim 300$ GeV, provides an increased phase space for parton emission 
and make multi-jet events possible.  
Connected to large $W$ is also the extension of Bjorken-$x$ down to 
$\sim 10^{-4}$ in DIS. New initial state radiation effects can then 
appear, such as BFKL-dynamics \cite{BFKL} where resummation of large 
$\log(1/x)$ terms may become noticeable in the structure function or the 
characteristic non-ordering in transverse momenta of emitted gluons observable
in forward energy flows or jets. 

Two main new observations have already been made in the small-$x$ data 
collected by H1 and ZEUS collaborations at HERA. The first is the large mean
transverse energy at forward rapidities \cite{H1-eflow,ZEUS-eflow} (i.e. in
the proton beam hemisphere), which may be a signal for BFKL dynamics. In
contrast to this mean behaviour,  the second effect is the discovery
\cite{ZEUS-gaps,H1-gaps} that about 10\% of  the events have no particles
or energy depositions in the forward part of the  detector, i.e. the final
state has a large gap in rapidity space.  These two features of the data are
thus `orthogonal' and hard to account  for in one and the same theoretical
framework. The description of the  energy flows requires a sufficient amount
of perturbative parton emission and non-perturbative effects from
hadronization. The gap events, on the other hand, require the absence of
parton and hadron emissions in the rapidity region of the gap.  A main
question is then whether these gaps can occur as fluctuations in perturbative
QCD and hadronization, with some new dynamical assumptions,  or whether their
understanding requires the addition of  a conceptually different 
mechanism such as diffractive interactions based on pomeron exchange
 where there is no continuous transition between the two phenomena. 

Given the complexity of the final states and these observables, a detailed 
comparison with theoretical models is best performed using Monte Carlo 
simulation methods. Standard programs, like {\sc Lepto} \cite{LEPTO61} and  
{\sc Ariadne} \cite{ARIADNE}, for DIS have been totally unable to
describe the  rapidity gap events. The gap events can, on the other hand, be
reasonably  well described in terms of diffractive interactions as given by
pomeron  exchange in Regge phenomenology. The basic idea is that the pomeron
has a  parton content which can be probed in hard scattering processes
\cite{IS}.  Monte Carlo models,  such as {\sc Pompyt} \cite{POMPYT} and {\sc
Rapgap} \cite{RAPGAP}, based on  these notions can give a reasonable
description of  the data. In spite of this, there is no satisfactory
understanding of the pomeron and its interaction mechanisms. It is therefore
of interest to consider alternative ways to understand the rapidity gap
phenomenon. 

We have previously introduced \cite{SCI} a novel soft colour interaction (SCI) 
mechanism to understand the large rapidity gap events in a new way, without 
any explicit use of the pomeron concept.
Our approach is similar in spirit to the one in ref.~\cite{BH}. Their model 
considers only the first order QCD matrix element, where the whole DIS 
cross-section is being saturated by the boson-gluon fusion process,
 and introduces a   
statistical probability that the produced $q\bar{q}$ pair is changed 
from a colour octet into a colour singlet state. 
In our model, the influence of higher order parton emissions are
included, an explicit mechanism for the colour exchange is introduced and
hadronization is taken into account. Having formulated our model in
terms of a Monte Carlo generator that simulates the complete final
state, more detailed comparisons with data can be made. 
Adding our SCI model to the normal Monte Carlo program 
{\sc Lepto} \cite{LEPTO65} for DIS, we describe the salient features of 
the HERA data, both regarding rapidity gaps and energy flows. 

In this paper we make a more 
comprehensive discussion of our model and describe its parts in more detail. 
The perturbative QCD aspects, i.e. matrix elements and parton showers,
are discussed in section 2. The non-perturbative part of the model, i.e. 
soft colour interactions, target remnant treatment and hadronization, 
are described in section 3. In section 4, we show the most relevant results 
in terms of observable quantities and make some comparison with data.
Finally, we give a concluding discussion in section~\ref{sec:conclusions}.

\section{Perturbative QCD effects}
\label{sec:pert}

The starting point for describing the hadronic final state is the deep
inelastic scattering on a quark in the proton as given by standard electroweak
cross  sections for $\gamma ,\: Z,\: W$ exchange \cite{LEPTO65}. Perturbative
QCD  (pQCD) effects are then taken into account through leading order  matrix
elements and higher orders through parton shower approximations.  These two
parts are discussed in the following, together with their  interrelation.

\subsection{Matrix elements}
\label{sec:me}

Hard parton processes are best described by exact matrix elements, but one  is
then limited to low orders in the perturbative expansion. The two leading
order QCD processes QCD Compton (QCD-C) and boson-gluon fusion (BGF) are 
shown in Fig.~\ref{fig:me_diagram}, which also defines the relevant
four-momenta. The two processes result in a quark-gluon and a quark-antiquark
parton system,  respectively, and are labeled by $qg$ and 
$q\bar{q}$.  The cross-section is calculated by folding the proton's parton
density functions (for which we use CTEQ2L \cite{CTEQ3})
with the hard scattering cross-sections
$d\hat{\sigma}_{qg}$ and $d\hat{\sigma}_{q\bar{q}}$ \cite{QCDME,KMS}
\begin{eqnarray} \label{eq:dsigma}
d\sigma_{ME}(x,Q^2,x_p,z_q,\phi) & \propto & \sum_{f}
  q_f(x/x_p,Q^2) \otimes d\hat{\sigma}_{qg}(x,Q^2,x_p,z_q,\phi) \nonumber \\
 & & + g(x/x_p,Q^2)  \otimes d\hat{\sigma}_{q\bar{q}}(x,Q^2,x_p,z_q,\phi).
\end{eqnarray}
$Q^2=-(p_l-p_{l'})^2$ and $x=Q^2/2P\cdot q$ are the usual deep inelastic 
variables that are sufficient to describe the inclusive lepton scattering. 
The three additional degrees of freedom in the first order processes are 
usually described in terms of $x_p=x/\xi$, $z_q=P \cdot j_1/P \cdot q$ and the 
azimuthal angle $\phi$ between the lepton scattering plane and the parton 
scattering plane.  The incoming parton momentum $k=\xi P$ is a fraction 
$\xi$ of the proton's momentum. The two produced partons have invariant mass 
squared $\hat{s}=(j_1+j_2)^2$, whereas the whole hadronic system has invariant
mass squared $W^2=(P+q)^2$.

\feynfig{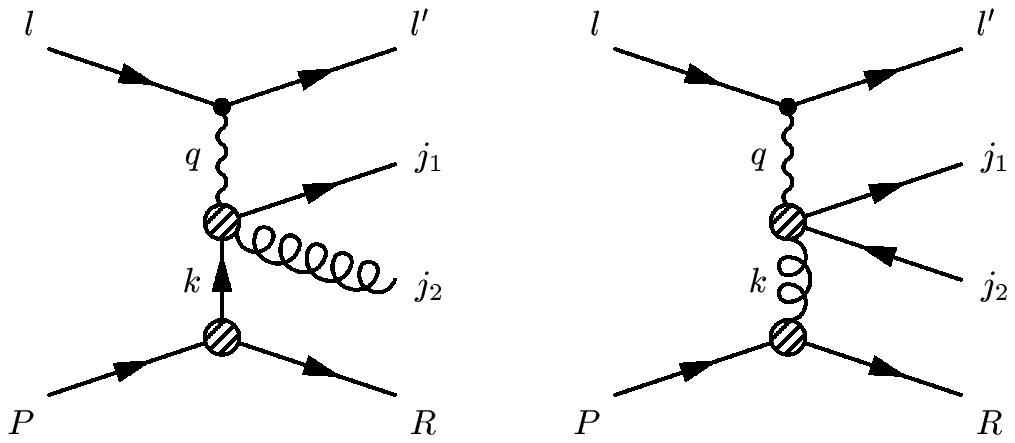}{4cm}{
First order QCD processes in DIS: (a) QCD Compton $\gamma^* q\to qg$, 
(b) boson-gluon fusion (BGF) $\gamma^* g\to q\bar{q}$. 
}{fig:me_diagram}

These matrix elements are often discussed only in connection with cross
sections for observable jets, corresponding to hard and well separated
partons.  Here we want to describe the hadronic system more generally and
therefore  also make extensions into the region where the 
hard subsystem has a small invariant mass squared $\hat{s}$.
One then has to care about the divergences in the matrix elements, which for
the QCD-Compton part $\hat{\sigma}_{qg}$ behaves as $1/(1-x_p)(1-z_q)$ and 
for the boson gluon fusion part  $\hat{\sigma}_{q\bar{q}}$ as $1/z_q(1-z_q)$
\cite{QCDME,KMS}.  In a Monte Carlo simulation, these are usually avoided
by a simple cut-off, although a more elaborate procedure
based on Sudakov form factors is also  possible \cite{Mike:ME-shower}. There
exist several different cut-off schemes  which have been considered~\cite{KMS}.
Most commonly used  has been the $W$-scheme with the requirement
$s_{ij}=(p_i+p_j)^2>y_{cut}W^2$ for any pair $ij$ of partons.  Here, not only
$j_1,j_2$ but also the remnant parton $R=(1-\xi)p$ is included 
to provide a cut against soft and collinear emissions relative to the initial
parton direction.  This scheme is essentially taken over from $e^+e^-$
annihilation and is  used in the JADE algorithm for jet reconstruction.

Here, we introduce the `$z\hat{s}$' scheme, which is a variant of the mixed 
scheme \cite{KMS}. In the $z\hat{s}$ scheme the soft and  collinear divergences
with respect  to the incoming parton direction are regulated with a cut in 
$z_q$ such that $z_{q,\min}<z_q<1-z_{q,\min}$. In addition there is also a cut
in the invariant mass of the hard subsystem, $\hat{s}>\hat{s}_{\min}$, which
regulates the soft and collinear divergences for the two produced partons. In
this way the  divergences with respect to the incoming parton direction and
with respect to the hard subsystem are treated separately. The advantages of
this scheme will become clear in the following.  

The difference between these two cut-off schemes is demonstrated in
Figs.~\ref{fig:me_xpzq} and \ref{fig:forward2}.  In the $W$ scheme the cut-off
in $z_q$ is to a good approximation given by $y_{cut}<z_q<1-y_{cut}$ whereas
the cut-off in $\hat{s}$ is given by $\hat{s}>y_{cut}W^2$. Since
$\hat{s}=Q^2(1/x_p-1)$ the cut in $\hat{s}$ can be translated into one in $x_p$
instead,  $x_p<Q^2/(\hat{s}_{\min}+Q^2)=x/(x+y_{cut})$. This means that if one
wants to use the matrix element for small $\hat{s}$ (i.e.~large $x_p$) one also
gets very close to the  divergences  at $z_q=0$ and $1$.  In the $z\hat{s}$
scheme on the other hand these divergency regions can be avoided and larger
$x_p$ can be reached since the two  divergences are handled separately. This
has the advantage that a region with a large 2+1-jet event rate can be treated
with the matrix element.  With smaller $\hat{s}$ one can, e.g., reach smaller
$x$ in an extraction of the gluon density $g(x,Q^2)$ from the boson-gluon
fusion process  \cite{Anders,H1-gluon}.

\ffigh{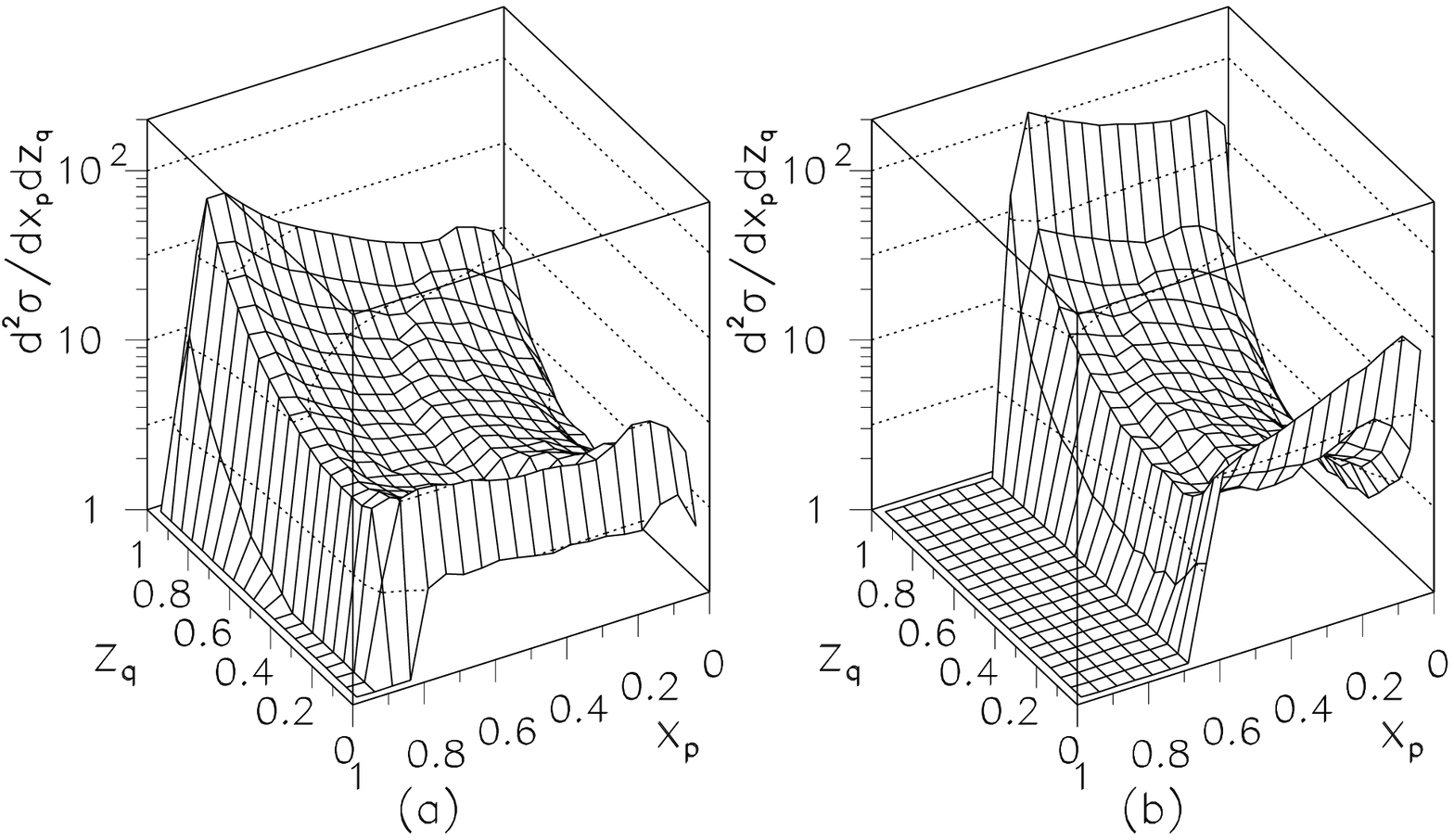}{8cm}{ Differential
first order QCD cross-section, eq.~(\ref{eq:dsigma}),  for
$x=10^{-3}$ and $Q^2=10$ GeV$^2$ in DIS at HERA using (a) the $z\hat{s}$
scheme with $z_{\min}=0.037$ and $\hat{s}_{\min}=1$ GeV$^2$, (b) the $W$ scheme
with $y_{cut}=0.00062$. The cuts have been chosen very small, such that  the
first order cross-section equals the total cross-section, to illustrate the
behaviour of the cross-section close to the divergences.}{fig:me_xpzq}

Fig.~\ref{fig:me_xpzq} illustrates the $2+1$ parton phase space for the two 
different schemes with the cut-offs chosen such that the $2+1$ parton cross
section equals the total cross-section. From the figure it is obvious
that the  $W$ scheme has the disadvantage to get very close to the
divergences  at $z_q=0$ and $1$, while the $z\hat{s}$ scheme avoids these
regions  and instead reaches larger $x_p$.

\ffigh{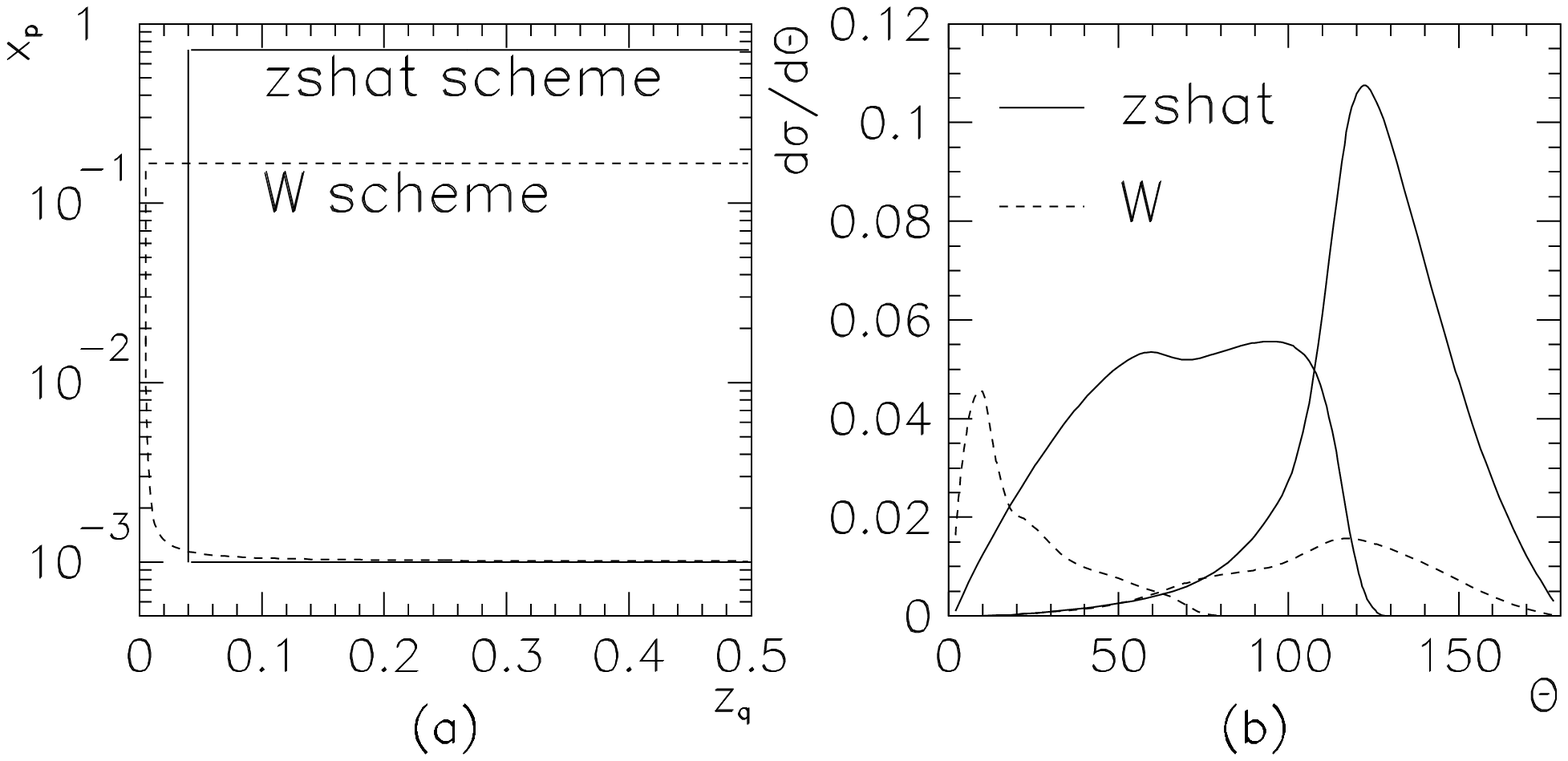}{7cm}
{ The 2+1 parton cross-section for $x=10^{-3}$ and $Q^2=10$ GeV$^2$ at HERA
using  the $z\hat{s}$ scheme (solid) with $z_{\min}=0.04$ and $\hat{s}_{\min}=4$
GeV$^2$ and the $W$ scheme (dashed) with $y_{cut}=0.005$. 
(a) The ($z_q,x_p$) phase space in the two schemes. Note that only the
region $z_q<0.5$ has  been shown due to the symmetry $z_q \leftrightarrow
1-z_q$ of the cut-offs.  
(b) Polar angle (with respect to the incoming proton) distribution 
for the most forward going parton of $j_1$ and $j_2$ and the other one
in the two schemes.}
{fig:forward2}

For the following studies we use the cut-offs $z_{\min}=0.04$ and
$\hat{s}_{\min}=4$ GeV$^2$ in the $z\hat{s}$  scheme and $y_{cut}=0.005$ in the
$W$ scheme, resulting in the phase-space region shown in
Fig.~\ref{fig:forward2}a. Once again this illustrates that one gets much closer
to the $z_q$ divergences in the $W$ scheme. The same cut-offs has also been
used to calculate the angular distributions of the two partons from the matrix
elements as  shown in Fig.~\ref{fig:forward2}b which demonstrates a clear
difference between the two schemes. In the $W$ scheme there is mostly one very
forward parton  (along the proton momentum direction) and one backward as
opposed to the  $z\hat{s}$ scheme which has much fewer forward partons. The
reach of smaller $\hat{s}$, in the latter scheme, means that the partons are
closer in phase space and thereby in angle. 

The fewer forward partons in the $z\hat{s}$ scheme, also imply that  forward
parton emission should be described by the initial state parton shower, as
described in next section. On the other hand, the $W$ scheme has larger
$\hat{s}$ and should have more  emissions in the final state parton shower.

Since we use lower cuts for the matrix element than in normal jet analyses,
where the jet cross-section is usually numerically a fraction $\alpha_s$ of the
total cross-section, we have to ensure that the jet cross-section is not
unphysically large. Large relative jet cross-sections also increase the risk of
double counting which for example would affect the energy flow. However, we
have explicitly  checked that increasing the $\hat{s}$ cut from $4$ GeV$^2$ to
$20$ GeV$^2$, which makes the jet cross-section be $\sim 20\%$ of the total,
does not influence the energy flows. Still, we want to use the smaller
$\hat{s}$ cut to improve the description of small mass diffractive systems with
the matrix element.

In the regions of small and large $z_q$ where the first order matrix elements
diverge rapidly one can expect large corrections from higher orders
\cite{MZdijet}. Therefore, parton emission in this region should be better 
treated with parton showers since they include the leading contributions from 
all orders (see next section). The fact that the $z\hat{s}$ scheme is rather
flat in the central region of Fig.~\ref{fig:me_xpzq}  corresponding to
the cut-offs used, shows that one is not getting too close to the
divergences.  The main principle should be to use the most appropriate
approximation,   finite order matrix element or leading log parton shower, in
each region  of the available phase space.

Thus, we use the first order matrix elements with the above  defined
$z\hat{s}$  cut-off scheme to describe the hard emissions and add parton
showers to take into account  higher order emissions. The procedure to
generate the perturbative  interactions in an event is then as follows. First,
the overall DIS  kinematical variables $x$ and $Q^2$ are chosen from the
electroweak cross-sections with QCD-improved structure function
\cite{LEPTO65}. Then,  the QCD matrix elements are used to choose whether to
generate a QCD-Compton or boson-gluon fusion process, or simply a quark-parton
model (QPM) quark  scattering without an extra parton emission from the matrix
elements,  i.e. a $qg$-, $q\bar{q}$- or $q$-event, respectively.  For $qg$-
and $q\bar{q}$-events, the matrix element is also used to choose  the
variables $x_p,\: z_q,\: \phi$ such that the parton four-momenta are  fully
specified \cite{LEPTO65}.  Parton showers are then added, as will now be
discussed.

\subsection{Parton showers}
\label{sec:shower}

Higher order parton emissions become very difficult to calculate using exact 
matrix elements due to the rapidly increasing number of diagrams. Second order
matrix elements have been calculated \cite{ME2}, but not implemented in any 
complete Monte Carlo program for event generation due to problems with a
probabilistic  interpretation. Higher orders are instead taken into
account through parton  showers based on the GLAP evolution equations
\cite{GLAP} in the  leading $\log Q^2$ approximation of perturbative QCD. 
Multiparton emission is thereby factorized into a series of emissions, each 
described by the basic branching processes $q\to qg$, $g\to gg$ and $g\to
q\bar{q}$, giving an iterative procedure suitable for Monte Carlo  simulation. 
The leading logarithm approximation implies that soft and collinear emissions 
should be well treated, whereas hard emissions at large angles are not and 
should instead be treated with matrix elements as far as possible. 

The parton showers is a way to take the off-shellness of the  incoming
and outgoing partons in the matrix element and the associated parton radiation
into account. The incoming parton, having a space-like virtuality due to the
radiation of time-like or on-shell partons in a quantum  fluctuation, can be
resolved by the hard scattering where the virtuality of the incoming parton is
limited by the hardness of the scattering.  The initial parton shower then
retraces the quantum fluctuation, which is realized by the scattering, back to
the originating parton in the incoming nucleon which has a space-like
virtuality below the cut-off  $Q_0^2 = 1$ GeV$^2$. The partons produced in
the hard scattering and by the  initial state parton shower may have time-like
virtualities ($m^2>0$) which is reduced in a final state parton shower by
branchings into daughter partons with decreasing off-shell masses and
decreasing opening angles. Finally all emitted partons have virtualities below
a cut-off $m_0^2 = 1$ GeV$^2$ and are put on-shell.
The separate treatment of initial and final state parton shower emissions 
implies the neglect of interference terms between the two and is not gauge 
invariant, but is consistent with the leading log approximation. 

\feynfig{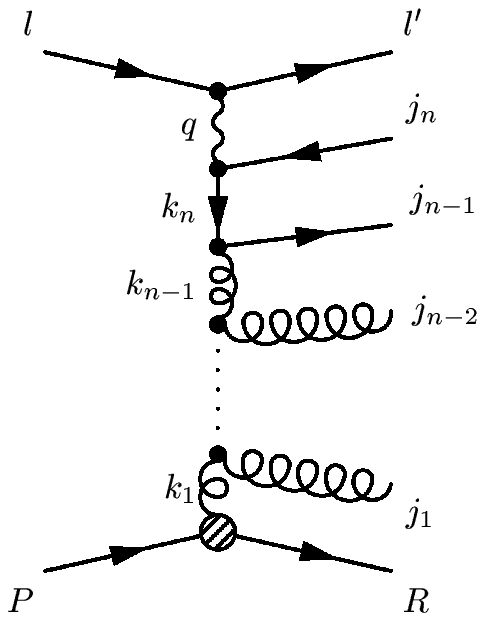}{6cm}{
Illustration of a DIS event with higher order parton emission in the parton 
shower approach. The emitted partons can have time-like virtualities which 
initiate final state parton showers.}{fig:shower_diagram}

The final state radiation is analogous to parton radiation in $e^+e^- \to
q\bar{q}$. The well developed and tested algorithm in {\sc Jetset} 
\cite{JETSET} is therefore used for all time-like showers in DIS. 
The technical details are given in \cite{PS}, but it should be noted that
coherence in soft gluon emission is taken into account through angular ordering
(decreasing opening angles in subsequent branches).

The initial state radiation is performed using the `backwards' evolution scheme
\cite{BACKWARD}. The shower is then constructed from the hard interaction, 
associated with the electroweak boson vertex, and evolved backwards with
decreasing  virtualities down to the essentially on-shell parton from the
incoming nucleon. This is a  more complicated process, since the nucleon parton
density functions must be taken into account in each step.
 When combining the initial and
final state radiation in DIS, special precautions must be taken to preserve
energy-momentum conservation and  keep the normal definitions of the $x,\: Q^2$
variables \cite{PS}.

The amount and hardness of the initial and final radiation depends on the
off-shellness of the partons entering and emerging from the hard scattering. 
These virtualities are chosen, using the Sudakov form factor, between the lower
cut-offs ($Q_0^2$ and $m_0^2$, respectively)  and a maximum regulated by the
hard momentum transfer scale. Having the parton showers as a higher order
correction to the matrix element  treatment implies adding them on a $q$, $qg$
or $q\bar{q}$ event. Emissions  harder than the ones treated by the matrix
elements  are then not allowed to avoid double counting.   Thus, in the case of
a $q$ event the maximum virtuality scale is set by the matrix element cut-off,
i.e.  $z_{\min}Q^2/x$ and $\hat{s}_{\min}$ for the initial and final state
parton showers, respectively, in the $z\hat{s}$ scheme or $y_{cut}W^2$ in the
$W$-scheme.  In the case of a $qg$ or $q\bar{q}$ event the maximum virtuality
scale  for the final state shower is set to $\hat{s}$.  For the initial state
shower, the natural scale  is the mass-squared of the parton propagator just
before the electroweak boson  vertex, cf. $|k_n^2|$ in
Fig.~\ref{fig:shower_diagram}. This is given by the known four-vectors of the
electroweak boson and  the two final partons from the matrix element.  
Depending on the underlying Feynman diagram in the amplitude,  different
combinations are possible leaving some remaining ambiguity and  the largest of
these possible virtualities has been chosen \cite{LEPTO65}.

Thus the scale for the showers are not simply given by $Q^2$, 
as may be naively expected based on GLAP evolution of structure functions.
Instead the relevant scales in the matrix element are used which
is quite in analogy with jet physics in photo-production
or in hadron-hadron scattering where the $p_\perp^2$ scale of the 
jets is used. This means that, at small $x$ in particular, the scale can be
significantly larger than $Q^2$.

The virtuality scale for the parton showers will vary from event to event, 
since they are regulated by the actual emission in the matrix element 
treatment. Or, expressed in another way, there is a dynamically changing  phase
space limit between the matrix element and the parton showers.   In each event,
all radiation softer than the matrix element emission is  given by the parton
showers. For $qg$- and $q\bar{q}$-events this means smaller $z_q$ and larger
$x_p$ in  Fig.~\ref{fig:forward2}a and for $q$-events the emissions are outside
the matrix-element region in the same figure. The typically smaller $\hat{s}$
in the $z\hat{s}$ scheme, leads to typically  less phase space for the final
state shower, but instead leave more phase space for the initial state shower. 
Ideally, the model should be stable against variations of this boundary 
between matrix element and parton shower treatments.  Since these are different
approximations in perturbative QCD there  will always be some residual
dependence on where the boundary between  them is drawn, however with the
cut-off scheme presently used the model is  quite stable when  the cut-offs are
varied. As emphasized before, the main principle should be to use the most
appropriate approximation in each region of the available phase space.

\subsection{Parton level results}
\label{sec:partonlevel}

Studies of emitted particles can preferentially be made in the 
hadronic cms. With the logarithmic variables rapidity 
and $\log p^2_\perp$, the phase space for parton emission 
has a triangular shape as  indicated
in Fig.~\ref{fig:triangle}. 
The density of partons from our Monte Carlo model is here shown and compared to 
the corresponding results of an alternative parton shower approach, 
the colour dipole model (CDM) \cite{CDM} as implemented in the {\sc Ariadne} 
Monte Carlo program \cite{ARIADNE}. The electroweak part is here the same, 
whereas the QCD parton emission description differs
(as well as some aspects of the non-perturbative proton remnant treatment). 

\ffigh{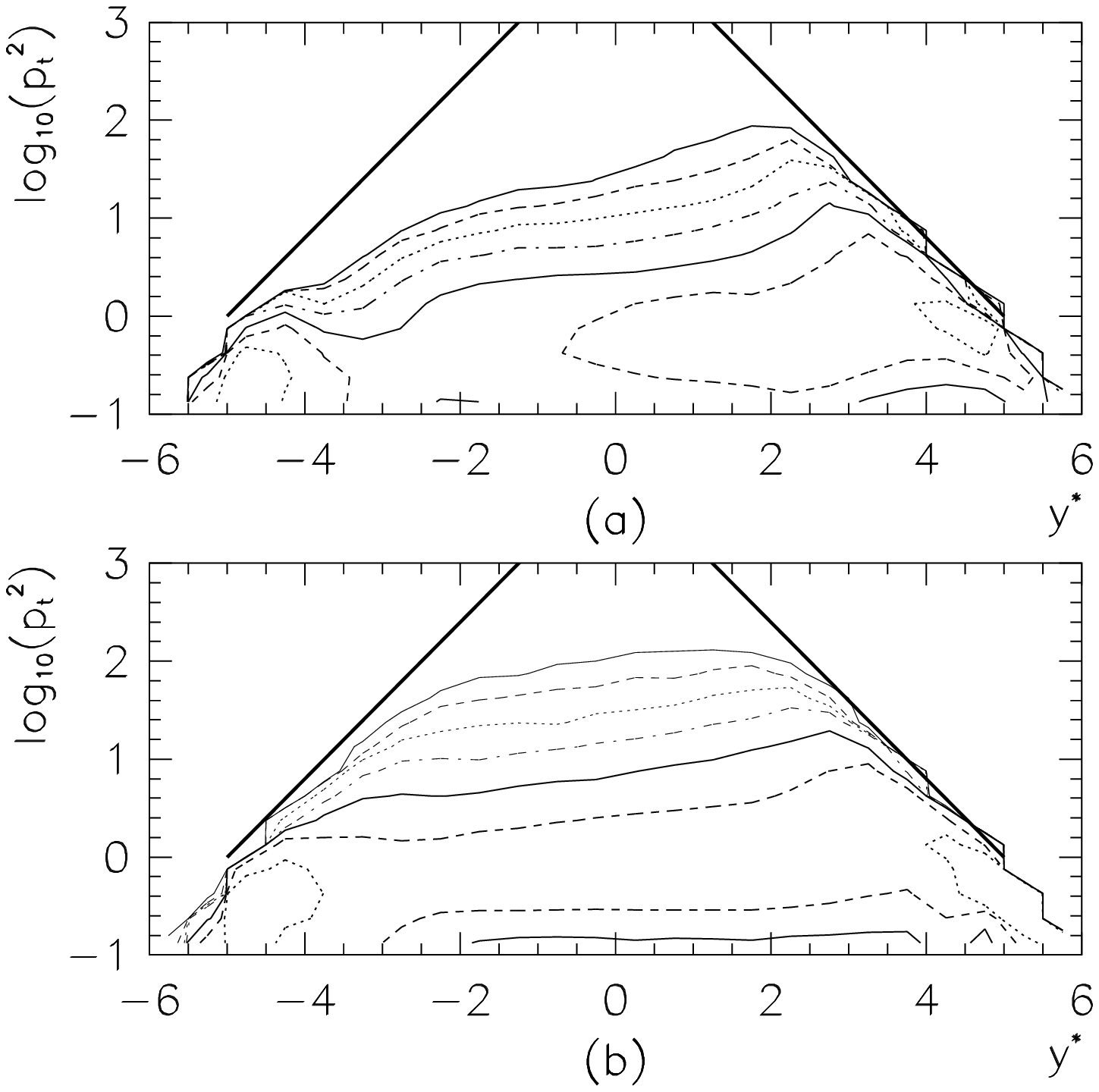}{85mm}{ Density
$dN/dy^* d\log_{10}(p^2_\perp)$ of emitted partons in terms of  their rapidity
$y^*$, with the proton remnant at $y^*<0$, and transverse momentum squared
$p^2_\perp$  in the hadronic cms as obtained from Monte Carlo events from (a)
{\sc Lepto}  and (b) {\sc Ariadne} for $x=10^{-3}$ and $Q^2=10$ GeV$^2$ at the
HERA energy.  The curves represent lines of constant density of emitted partons
with a  factor two change between adjacent curves and the same scale in both
figures.  The thick outer lines
indicate the triangular phase space boundary.}{fig:triangle}

In the CDM model, the partons from the hard scattering together with
quarks and diquarks in the proton remnant form colour dipoles  which radiate
additional partons, such that new dipoles are formed and the  radiation
procedure is iterated. Formally, there is no initial state radiation, but the
dipoles radiate in the phase space region normally associated with  initial
state radiation and therefore effectively simulates both initial and  final
state radiation. Parton densities are not taken into account, 
as in the normal GLAP
radiation, and there is therefore no related suppression of the `initial state' 
radiation.  Instead, the CDM model has a suppression 
due to the extended nature of the proton remnant. 

Another important difference between the GLAP-based parton emission scheme 
and the CDM one, concerns the $p_\perp$ ordering of the emitted partons. 
The GLAP formalism is based on a strong ordering in virtuality of the parton 
propagators ($k_i$ in Fig.~\ref{fig:shower_diagram}) and thereby of the 
transverse momenta of emitted partons ($j_i$). This requirement is not present
in the CDM model, where the transverse momenta may fluctuate as one is tracing
the emissions in rapidity. 

Comparing the GLAP-based and CDM-based Monte Carlo results in 
Fig.~\ref{fig:triangle}ab, one can clearly see the expected difference.  The
radiation in the {\sc Lepto} model is more suppressed the closer to the proton 
remnant one gets, whereas the {\sc Ariadne} model gives a population of
partons which extends more towards the proton remnant. 
In fact there is only little suppression close to the proton 
remnant in {\sc Ariadne}, such that the radiation is almost symmetric between
the proton remnant and the scattered quark. The suppression in
{\sc Lepto} is partly due to the virtuality ordering in GLAP and
partly to the effect of taking the proton's parton  densities into account in
the initial state evolution.  
This suppression is an essential feature of the
radiation since the soft colour interactions introduced later cannot give
rapidity gaps unless there are gaps already at parton level.
The comment in \cite{Leif-gaps} that soft parton radiation should
destroy possible gaps depends on the assumption that there is hardly no 
suppression of the radiation in the dipole model. 
Taking the parton densities into account the probability will increase for
a gap at the parton level.

\section{Non perturbative effects}
\label{sec:nonpert}

To make the transition from the perturbative parton level interactions
discussed so far, to the final state of observable hadrons various
non-perturbative  effects have to be taken into account. This can be achieved
by different  phenomenological hadronization models. Such models are not
based on  first principles or derived from the QCD Lagrangian and therefore
there are uncertainties and variations possible even in models that are
well-functioning in describing many experimental observations. 

In the following we discuss two aspects of non-perturbative effects of 
relevance for DIS. First, the treatment of a target proton remnant containing
a sea quark in addition to the valence quarks. Secondly, the newly introduced
\cite{SCI} hypothesis of soft colour interactions which may change the colour
structure and thereby produce rapidity gap events. 

\subsection{Proton remnant with a sea quark}

The remnant system is the target nucleon `minus' the parton entering 
the hard scattering system (initial parton showers and matrix
elements). This interacting parton can be either a valence quark, 
a sea-quark or a gluon. In case a valence quark is removed the remainder is
a  diquark which is taken as a colour anti-triplet at the end of a
string, to which Lund model hadronization is applied. However, if a sea
quark is  removed the remainder is more complex, with the valence quarks
plus  the partner of the interacting sea quark to conserve 
quantum numbers.  Here, we will only discuss the physics aspects of this 
latter case, where an improved treatment has been introduced \cite{SCI}.
For the other cases and technical details we refer to \cite{LEPTO65}.

The main idea in the new sea quark treatment is that the removed quark is 
assigned to be a valence or sea quark and, in case of a sea quark,  
its left-over partner in the remnant is given some dynamics. 
Thus, the interacting quark is taken as a valence or sea quark from the
relative sizes of the corresponding parton distributions
$q_{val}(x_1,Q^2_1)$ and $q_{sea}(x_1,Q^2_1)$ 
where $x_1$ is the known
momentum fraction of the quark `leaving' the proton and $Q^2_1$ is the
relevant scale (typically the cutoff $Q_0^2$ of the initial state
parton shower). In case of a sea quark, the left-over partner is given
a longitudinal momentum fraction from the
Altarelli-Parisi splitting function $P(g\to q\bar{q})$ 
and the transverse momentum follows from the kinematics once the
partner mass has been fixed. Essentially the same results are obtained if the
longitudinal momentum fraction is chosen from the corresponding sea quark 
momentum distribution. The former approach is presently used since this
allows the mechanism to be simply implemented in the initial state
parton shower routine as an additional, but non-perturbative, $g\to
q\bar{q}$ process. 

\ffigh{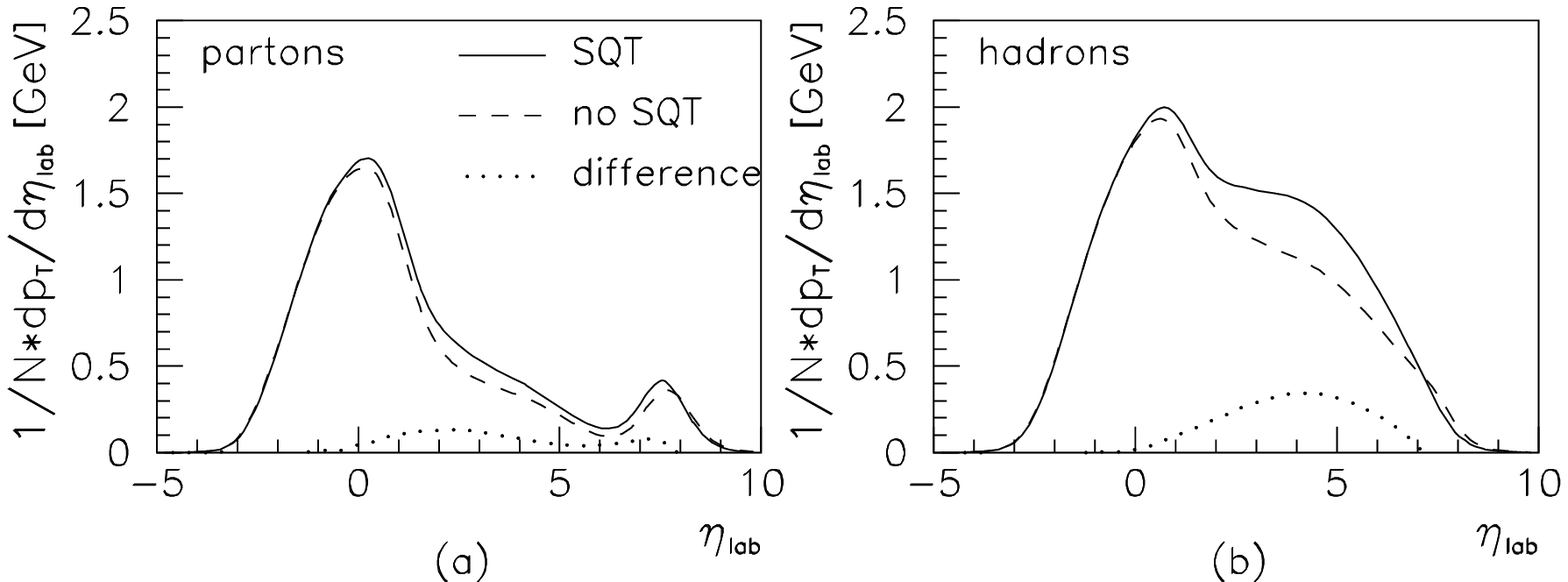}{55mm}{
Transverse energy flow at (a) parton and (b) hadron level from {\sc Lepto}
with the new target remnant sea quark treatment (SQT, solid curves) and without
(dashed curves), as well as their difference (dotted curves); based on 
simulated HERA events within $0.03<y<0.7$ and $7.0 < Q^2 < 70$ GeV$^2$. 
}{fig:tflow}

The sea quark partner and the three valence quarks in the remnant, which are
split into a quark and a diquark in the conventional way  \cite{LEPTO65}, form
two colour singlet systems (strings)   together with the sea quark partner and
the interacting sea quark (and any emitted gluons). This two-string
configuration for sea quark initiated processes provides a desirable continuity
to two-string gluon-induced events, arising  from the boson-gluon fusion matrix
element or a $g\to q\bar{q}$ splitting  in the initial state shower. This
reduces the dependence on the cut-off values in the matrix element and the
parton shower. Depending on the partner sea quark momentum, the corresponding
string will extend more or less into the central region in rapidity. The
hadronization of this extra string will thereby  contribute to the forward
energy flow, as illustrated  in Fig.~\ref{fig:tflow}. By comparing the effects
of the sea quark treatment on parton and hadron level it is clear that the main
effect is due to the extra string between the sea quark partner and the
remnant, and not so much from the sea quark partner itself.

A more general conclusion can also be drawn from Fig.~\ref{fig:tflow}, namely
that the forward energy flow is strongly influenced by hadronization in our
model.  This is obvious from the large difference between the parton level
and hadron level results. It is therefore non-trivial to use this observable
as  a test of detailed perturbative dynamics, e.g. to discriminate between
BFKL  and GLAP behaviour.  

\ffigh{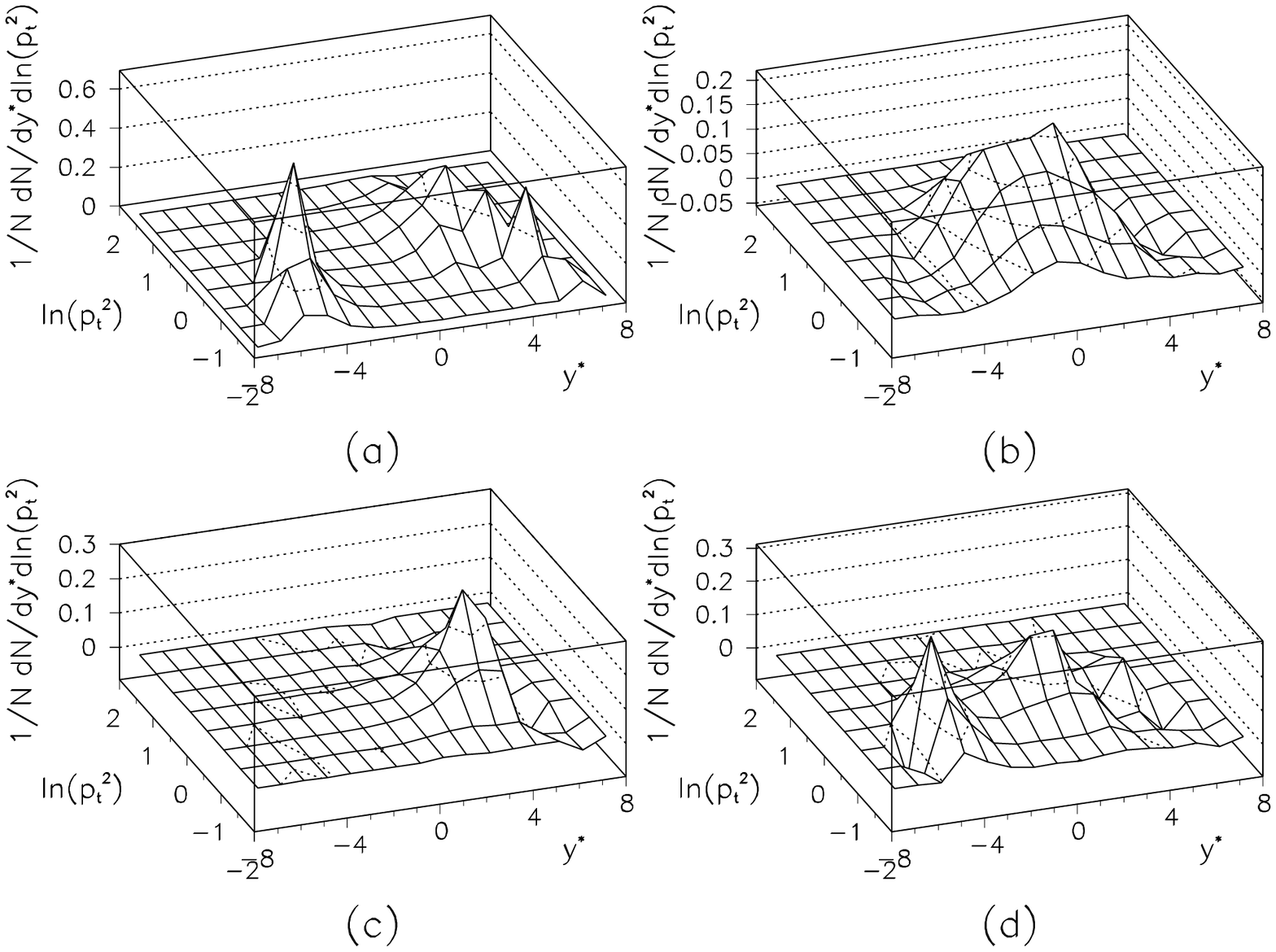}{11cm}{
Distribution in rapidity and transverse momentum (in the hadronic cms) 
for different emissions in {\sc Lepto}: 
(a) partons from the matrix element (right peak) and in the remnant (left peak),  
(b) partons from the initial state shower, 
(c) partons from the final state shower, 
(d) remnant partons including partner sea quarks.
In all cases obtained from simulated HERA events with $x=10^{-3}$ and
$Q^2=10$ GeV$^2$.(a) is from a simulation with only the matrix element,
while (b)-(d) are the difference of a full simulation and one with the examined
effect turned off.
}{fig:shower_ps}

To disentangle the effects of different processes producing 
partons we turn each of them off and subtract the result from the total 
simulation and demonstrate the result in Fig.~\ref{fig:shower_ps} in terms 
of rapidity and transverse momentum in the hadronic final state. 
Fig.~\ref{fig:shower_ps}a demonstrates that partons from the matrix elements 
are mainly in the current hemisphere ($y^*>0$) 
and have perturbatively large $p_\perp$, 
whereas the remnant partons are at large rapidity in the proton direction 
and have small $p_\perp$, as expected. Partons from the initial state shower 
(Fig.~\ref{fig:shower_ps}b) are distributed as a `ridge' in rapidity, whereas 
partons from the final state shower (Fig.~\ref{fig:shower_ps}c) are more 
concentrated in the current region. The remnant partons 
(Fig.~\ref{fig:shower_ps}d) are mainly at large negative rapidities, but the
location of the sea quark partner extends to central rapidities in the hadronic 
cms.  
Considering the absolute normalization on the vertical scales, one can note 
that the main contributions come from the matrix element and the valence 
partons in the remnant, followed by the initial state parton shower. 
The relative amount and distribution from matrix element and parton showers 
does to some extent depend on the cut-off procedure, as discussed before. 
 One should also note that there are negative contributions from the showers
and the sea quark treatment. This is just a consequence of conservation of
energy. If more particles are produced centrally with non-negligible transverse
masses then the `outer' particles have to be more central to preserve the
total hadronic energy, $W$.

\subsection{Soft colour interactions}

The starting point for hadronization is the partonic event generated by the 
matrix element and the parton showers. This implies a colour separation 
between the hard scattering system and the proton remnant, which will be 
connected by colour flux tubes, or strings in the Lund model \cite{Lund}. 
As a result of hadronization, the whole rapidity region in between will be 
populated by hadrons and the probability for a region with no particles is
exponentially suppressed with increasing `gap' size. 

The discovery of a rather large fraction ($\sim 10\%$) of DIS events at HERA
with a large gap in the distribution of final state particles is therefore
quite remarkable \cite{ZEUS-gaps,H1-gaps}.  These events cannot be understood
by conventional DIS models as implemented in  {\sc Lepto} \cite{LEPTO61} and
{\sc Ariadne} \cite{ARIADNE}.  They can be reasonably well described by
phenomenological models based on  pomeron exchange \cite{POMPYT,RAPGAP}, but
there is no satisfactory understanding of  the pomeron and its interaction
mechanism. As an alternative, we have therefore introduced \cite{SCI} the
concept of soft colour interactions (SCI) that give  rise to rapidity gap
events without using a pomeron from Regge phenomenology. 

\ffigh{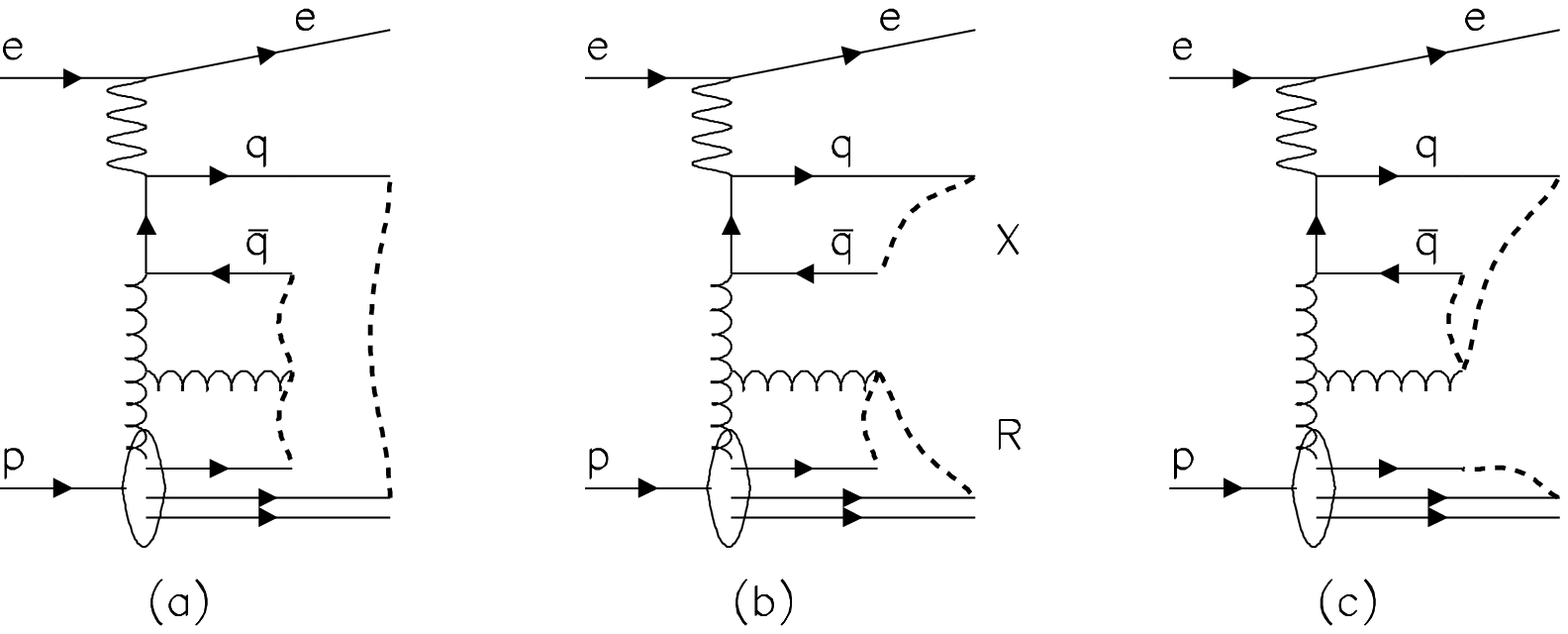}{50mm}{
The string configuration in a gluon-initiated DIS event in 
(a) conventional Lund string model connection of partons, and  
(b,c) after reconnection due to soft colour interactions.
}{fig:string}

The basic idea is here that there may be additional interactions between the 
partons at a scale below the cut-off $Q_0^2$ for the perturbative treatment. 
Obviously, interactions will not disappear below this cut-off, the question is 
how to describe them properly. Our proposed SCI mechanism can be 
viewed as the perturbatively produced quarks and gluons interacting softly 
with the colour medium of the proton as they propagate through it. 
This should be a natural part of the process in which `bare' 
perturbative partons are `dressed'  into non-perturbative ones 
and the formation of the confining colour flux tube in between them. 
These soft non-perturbative interactions cannot change the momenta of 
the partons significantly, but they may change their colour and thereby 
affect the colour structure of the event. This corresponds to a modified 
topology of the string in the Lund model approach, as illustrated in 
Fig.~\ref{fig:string}. 

Lacking a proper understanding of non-perturbative QCD processes, we construct 
a simple model to describe and simulate such soft colour interactions. All
partons from the hard interaction plus the remaining quarks in the proton
remnant, including a possible sea quark partner, constitute a set of colour
charges.  Each pair of charges can make a soft interaction changing only the
colour and not the momentum,  which may be viewed as soft non-perturbative gluon
exchange.  As the process is non-perturbative the exchange probability for a
pair  cannot be calculated so instead we describe it by a phenomenological  
parameter $R$.

The number of soft exchanges will vary event-by-event and change  the colour
topology of the events such that, in some cases, colour singlet subsystems
arise separated in rapidity.  In the Lund model this corresponds to a modified
string stretching as illustrated in Figs.~\ref{fig:string}bc,  where (b) can be
seen as a switch of anticolour between the antiquark and the diquark and (c) as
a switch of colour between the two quarks.  This kind of colour switches
between the perturbatively produced partons and the partons in  the proton
remnant are of particular importance for the gap formation.

As a result of soft colour interactions, there may be colour singlet systems 
with so small invariant mass (cf.~Fig.~\ref{fig:string}c)
that normal hadronization models are not 
applicable, since the final state is very constrained and resonance effects 
are important. This applies in particular to string systems including a
valence diquark from the remnant. Such systems are not optimally treated in 
 {\sc Jetset} regarding the production of one- and 
two-particle final states and taking isospin constraints into account.  For
instance, a $\Delta$-resonance is often made instead of a  nucleon-pion
system.  We have therefore constructed a new treatment of such systems in
order to account for effective isospin singlet exchange to prevent the
production of a  single  $\Delta$ resonance and instead make two particles if
kinematically allowed.

Our procedure is similar to the original one in {\sc Jetset}, but differs in 
details which, however, are important for the results. 
First, all particles in the system are added into a cluster.
If the invariant mass of the cluster is large enough 
then the cluster is decayed into two hadrons
isotropically (i.e.~according to $d\!\cos\!\theta \, d\phi$) in the 
centre-of-mass system of the cluster.
Just as in ordinary string hadronization the flavours in the 
cluster break are chosen from the $SU(6)$ weights which gives the two
hadrons to be formed.

If the cluster mass is too small to make two hadrons, a single hadron is made.
The type of hadron to be produced is chosen from the quark and diquark 
flavours using the $SU(6)$ weights, but applying the mentioned isospin 
constraint (against $\Delta$-resonances). 
The invariant mass of the cluster is generally different from
the mass of the produced hadron, typically a proton or a neutron,
and hence energy and momentum cannot be conserved unless the four-momentum 
of another parton in the event is also changed. 
To disturb the parton state as little as possible, 
the parton $p_i$ is chosen which together with the cluster momentum $p_c$ 
gives the largest invariant mass $(p_i+p_c)^2$. In this way the relative
changes needed in the four-momenta are very small 
(typically $\lsim 10^{-3}$ at HERA).
New four-momenta are then assigned according to
$p_c^\prime = (1+\varepsilon_1)p_c-\varepsilon_2p_i$ and 
$p_i^\prime = (1+\varepsilon_2)p_i-\varepsilon_1p_c$, 
with the conditions that the invariant mass of the other parton is not changed, 
${p_i^\prime}^2 = p_i^2$, and that the produced hadron 
gets the  correct mass, ${p_c^\prime}^2 = m_h^2$.

In systems of small mass there is a sensitivity on the assumed masses for 
a quark ($m_q$) and a diquark ($m_{qq}$), as well as the transverse 
momentum $s_\perp$ introduced in the target remnant split \cite{LEPTO65}. 
In order not to have a cluster mass larger than 
the proton mass just from such effects, the quark masses used in {\sc Jetset}
are lowered from 325 to 225 MeV. 
This reduction has been chosen such that $m_q+m_{qq}+\sigma_s \sim m_p $, 
where the Gaussian width $\sigma_s=\sqrt{2 \langle s_\perp^2 \rangle}=0.30$ 
GeV has been chosen from comparison with EMC data on target remnant protons. 
This, and the resulting properties of the forward-moving $R$-system is 
discussed in next section. 

\section{Observables and data}
\label{sec:observables}

\subsection{Remnant characteristics}

The model's treatment of the target remnant can be tested against data 
on baryon production in the target fragmentation region as obtained in 
muon-proton scattering by the European Muon Collaboration (EMC). In the
analysis of ref.~\cite{Erdman}, the data on proton and $\Lambda$ 
production have been analyzed in terms of $\Delta y =
\ln[({E+|\vec{p}|})/{m_B}]$, which is the difference in rapidity
between the target proton and the outgoing  baryon $B$ and corresponds
to having the $z$-axis along the $B$ momentum. The difference of
baryons and antibaryons have been taken, to cancel the  contribution
from baryon-antibaryon pair production in the hadronization  and the
results are shown in Fig.~\ref{fig:emc}. 

\ffigh{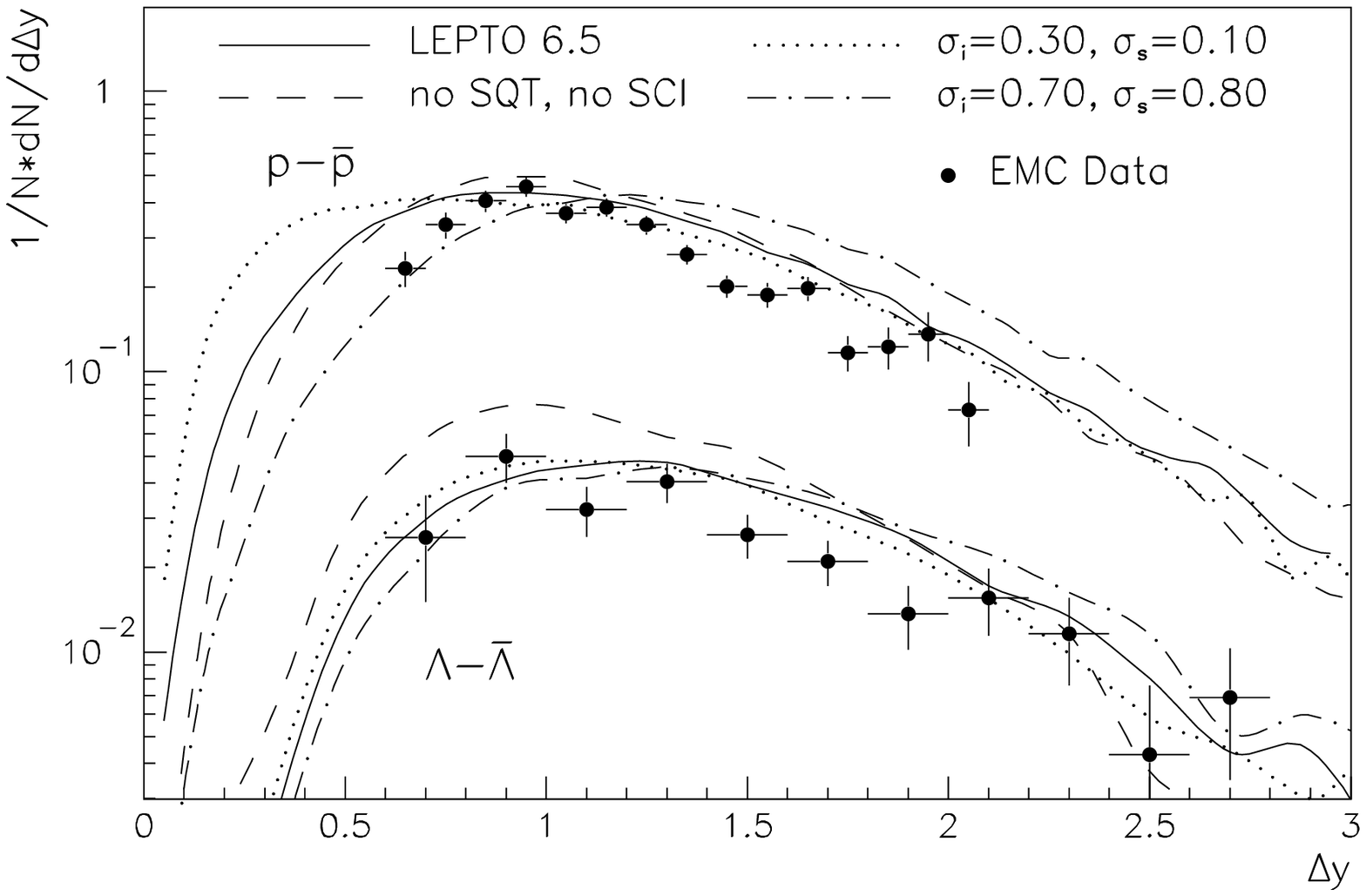}{8cm}{
Distribution in rapidity difference, $\Delta y = \ln[(E+|\vec{p}|)/m_B]$,  of
produced baryons relative to the target proton. The difference 
baryon--antibaryon enhances the effect of target fragmentation relative to 
pair production.  Data points from EMC and curves from the complete new model
with standard parameters (solid) and with varied Gaussian transverse momentum 
widths $\sigma_i$ and $\sigma_s$ for intrinsic momentum and target remnant
split,  as indicated, and the old model without SCI and sea quark treatment
(SQT).  
}{fig:emc}

The proton spectrum can be
well  described with both the old and the new models, whereas the
$\Lambda$ spectrum is better reproduced with the new model including
the improved sea quark treatment and soft colour interactions. 
In the old sea quark treatment, 
$\Lambda$ particles were produced directly from the proton remnant 
and not from a string hadronization which gave a too hard 
$\Lambda$ spectrum (small $\Delta y$).
As illustrated, the data also gives  some handle on the width $\sigma_i$ of
the intrinsic transverse momentum  associated with the parton's Fermi
motion in the target, as well as the  Gaussian width $\sigma_s$ of the
transverse momentum introduced in a target  remnant split. This gives
further motivation for the chosen values for these  parameters, i.e.
$\sigma_i=0.44$ GeV and $\sigma_s=0.30$ GeV. 

\subsection{Diffraction and gap events}
 
The amount of rapidity gaps will depend on the parameter $R$, but as was shown
in \cite{SCI} the dependence is not strong. In fact, increasing $R$ above
$0.5$, which we use, does not give an increased gap probability, but may
actually decrease it depending on the details of the colour exchanges in the
model. This is intuitively understandable, since once a colour exchange with
the spectator has occured additional exchanges among the partons need not
favour gaps and may even reduce them. 

In order to define the $X$ and $R$ systems and compare with experimental
data  we have used a gap definition similar to \cite{H1-gaps}, i.e. no energy
in  $\eta_{\max}<\eta<6.6$ with $\eta_{\max}<3.2$. (With $\eta
=-\ln{\tan{\theta/2}}$ and $\theta$ the angle relative to the proton beam so
that $\eta > 0$ is the  proton hemisphere in the HERA lab frame.) The main
features of the resulting  $R$-system in our model are shown in
Fig.~\ref{fig:remn}.  Here, $t=(p_p-p_R)^2$ is the momentum transfer from the
incoming proton to the  emerging $R$-system, $M_R$ is the mass of the
remainder system defined as all particles with $\eta>6.6$, 
$x_L=p_z/E_p$ is the longitudinal  momentum fraction 
of final state protons. 

\ffigh{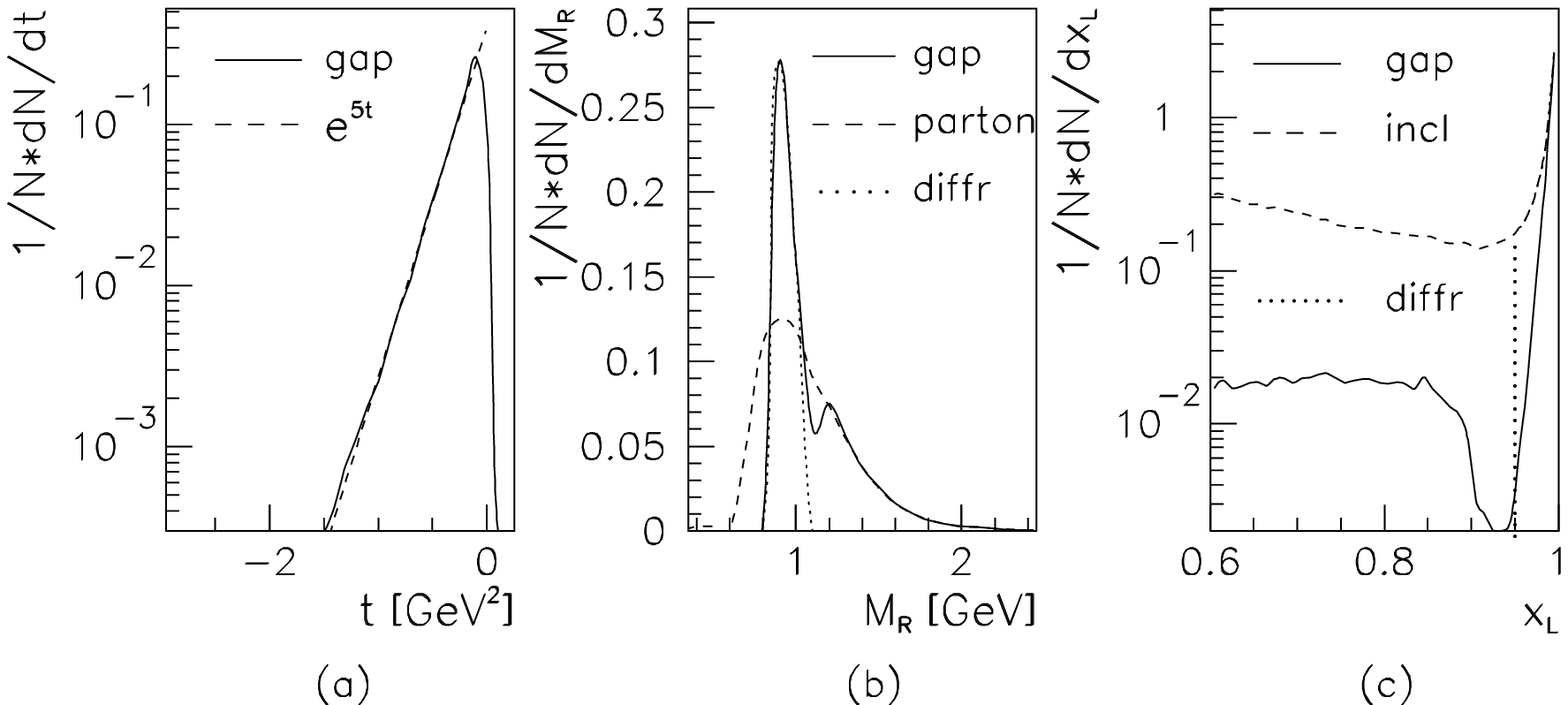}{70mm}{ (a) Squared
momentum transfer $t$ from incoming proton to remnant system $R$  for `gap'
events compared with the exponential slope $1/\sigma_i^2=5$ GeV$^{-2 }$. (b)
Invariant mass $M_R$ of the forward remnant system for `gap' events at the
hadron level (solid line) and parton level (dashed line) compared with
`diffractive' events at hadron level (dotted line). (c) Proton momentum
distribution for `gap' events (solid line), all events (dashed line) and
`diffractive' events (dotted line).}{fig:remn}

The $t$-dependence in our model is intimately connected to the assumed 
distribution of intrinsic transverse momentum ($k_{1\bot}$ in
Fig.~\ref{fig:shower_diagram}) of partons in
the proton, i.e.\ of the parton entering the hard scattering process.
This transverse momentum is balanced by the proton remnant and, since 
momentum transfers in SCI are neglected, it is essentially the $p_\bot$
of the forward $R$-system, i.e.\ $p_{R,\bot}^2=k_{\bot}^2$. Now,
$t\approx -p_{R,\bot}^2$ in the case of interest, i.e.\ when the 
energy-momentum transfer from the beam proton to the $X$-system is
small giving a very forward $R$-system. 

The intrinsic, or primordial, $k_\bot$ represents the non-perturbative Fermi 
motion in the
proton and is therefore of the order $k_{\bot}\simeq 1 \: \mbox{fm}^{-1}$ or a
few hundred $MeV$ as estimated from the uncertainty principle. This gives the
width $\sigma_i$ of the Gaussian distribution
$\exp{(-k_{\bot}^2/\sigma_i^2)}dk_{\bot}^2$ which is normally assumed.  
Thus, one
directly gets the exponential $t$-dependence $\exp(t/\sigma_i^2)dt$ with
$\sigma_i^2=2\langle k_{\bot}^2 \rangle$ from the primordial
$k_\bot^2$-distribution. The default value $\sigma_i=0.44$ GeV, obtained 
from fits to normal DIS data, then gives an effective  
diffractive-like $e^{5t}$ distribution as demonstrated in Fig.~\ref{fig:remn}a. 

The invariant mass of the forward $R$-system is shown in Fig.~\ref{fig:remn}b.
Here, one should first note that the mass spectrum for the partons in the
target  remnant peaks around the proton mass, i.e. $\sqrt{(p_q+p_{dq})^2}\sim
m_p$.  The $R$-system is therefore dominantly a single proton, as in a
diffractive  model based on pomeron exchange, but there is also a tail
corresponding to the proton remnant dissociating into a larger system. The
details of the  spectrum depends, however, on the above discussed treatment of
small mass systems.

Fig.~\ref{fig:remn}c shows the longitudinal momentum spectrum of protons in our
model and demonstrate a clear peak at large $x_L$.  Based on this distribution
we define events having a leading proton with  $x_L>0.95$ as `diffractive'. As
demonstrated in the figure, most but not all  of these events fulfill the above
gap definition.  In addition to the diffractive peak there are also gap events
with protons that have $x_L<0.95$. In these events the remnant system typically
consists of a nucleon and a pion, cf.~the difference of the solid and dotted
lines in Fig.~\ref{fig:remn}b. Fig.~\ref{fig:remn}c also shows that there is a
clear distinction between the two cases.

Rapidity gaps have been experimentally investigated \cite{ZEUS-gaps,H1-gaps}
through the observable $\eta_{\max}$ giving, in each event, the maximum 
lab pseudo-rapidity where an energy deposition is observed.  
Fig.~\ref{fig:xsyst} shows the distribution of this quantity as
obtained from our model simulations for $7.0<Q^2<70$ and $0.03<y<0.7$,
corresponding to the experimental conditions. In addition, the limit
$\eta<3.65$ in the H1 detector is taken into account. Clearly,
the introduction of soft colour interactions have a large
effect on the $\eta_{\max}$ distribution. Still, our SCI model is not
very sensitive to the exact value of the colour exchange probability 
parameter $R$ as was shown in \cite{SCI}. 

\ffigh{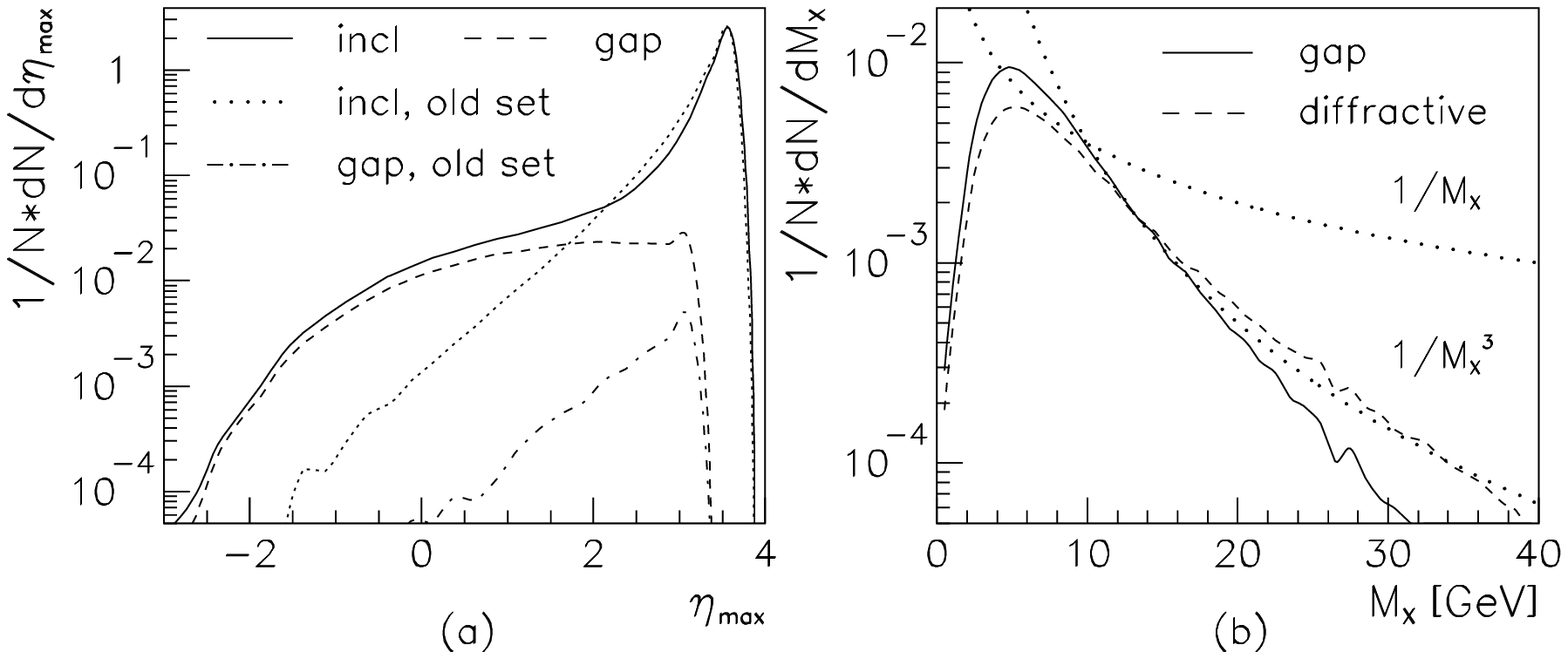}{70mm}{(a) 
Distribution of maximum pseudorapidity 
($\eta_{\max}$). Hadron level after soft colour interactions
for all events (solid line) and those satisfying the `gap' definition 
(dashed line). The case with no colour reconnections and sea quark
treatment is also shown for all events (dotted line) and `gap' events
(dash-dotted line). (b) Invariant mass of
the produced central system $X$ for `gap' events (solid line)
and `diffractive' events (dashed line) compared with
a $1/M_X$- and a $1/M_X^3$-distribution.}{fig:xsyst}

One should note that the basic features of this distribution, the
height of the peak and the `plateau', is in reasonable agreement with
the data \cite{ZEUS-gaps,H1-gaps}. A direct comparison requires a detailed
account of experimental conditions, such as acceptance and varying
event vertex position. Selecting events with rapidity gaps similar to
the H1 definition (i.e.\ no energy in $\eta_{\max}<\eta<6.6$ where
$\eta_{\max}<3.2$) gives the dashed curve in Fig.~\ref{fig:xsyst}, 
also in basic agreement with data \cite{H1-gaps}.
The drastic effect of the soft colour interactions on the $\eta_{\max}$ 
distribution is clearly demonstrated by comparison to the case when they 
are switched off, which is also included in Fig.~\ref{fig:xsyst}a. 

The $M_X$-dependence in Fig.~\ref{fig:xsyst}b is close to the 
$dM_X/M_X^3$ or $ds_{q\bar{q}}/s_{q\bar{q}}^2$ behaviour of the BGF
matrix element  with the cut-off against divergences visible at small
$M_X$. The large-$M_X$ region is distorted by the requirement of a gap
extending into the central rapidity region. Kinematically, larger $M_X$
means a reduced gap and is therefore disfavoured by the gap condition. 
This can also be seen from the $M_X$-distribution for  diffractive
events where there is no depletion for large $M_X$ and instead the 
distribution follows $dM_X/M_X^3$.

\subsection{The diffractive structure function}

To compare our model with experimental data on rapidity gap events we
consider the diffractive structure function $F_2^D(\beta,x_\pom,Q^2)$
defined by \cite{IP}
\begin{equation}
\frac{d\sigma^{D}}{d\beta dx_\pom dQ^2}=
\frac{4\pi\alpha^2}{\beta Q^4}\left[1-y+\frac{y^2}{2}\right]
F_2^D(\beta,x_\pom,Q^2)
\end{equation}
(i.e. assuming single photon exchange and neglecting $F_L$). This
inclusive quantity contains the dependence on the main variables 
$\beta \simeq Q^2/(Q^2+M_X^2)$, $x_\pom \simeq (Q^2+M_X^2)/(Q^2+W^2)$ 
and the usual DIS momentum transfer squared $Q^2$. The acceptance
corrected data from H1 \cite{H1-gaps} are compared in Fig.~\ref{fig:f2d3} 
to our model
results obtained by selecting Monte Carlo events with rapidity gaps similar to
the H1 definition (i.e.\ no energy in $\eta_{\max}<\eta<6.6$ where
$\eta_{\max}<3.2$).

The model is generally in good agreement with the data. It has a tendency to be
below the data at large $Q^2$, possibly due to slightly too much parton
radiation \cite{SCI}.  The $\beta$-dependence seems to be the same in the model
and the  data and thereby the $M_X$ dependence is also basically correct. At
large $\beta$ there is a slight tendency for the model to be below the data.
This is natural since large $\beta$ corresponds to small $M_X$ which are not
include in the model  ($\hat{s}_{\min}=4$ GeV$^2$) since it is outside the
perturbative  regime.

\ffigh{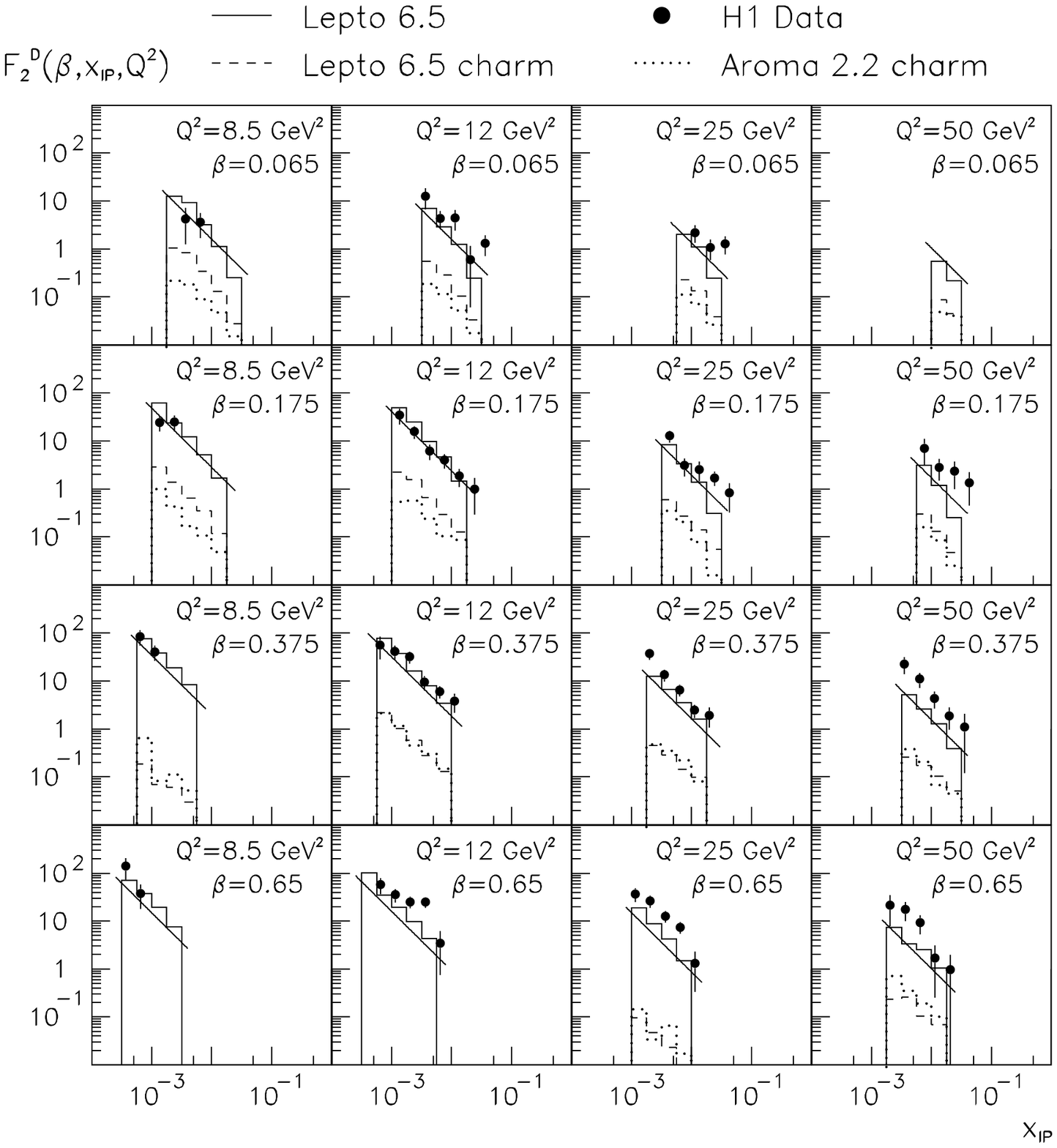}{150mm}
{The diffractive structure function  $F_2^D(\beta,x_\pom,Q^2)$  from
our soft colour interaction model using the `gap' definition (solid histograms)
and the `diffractive' definition (straight lines) compared to  the H1 data
points \protect\cite{H1-gaps}. Also shown are the charm contributions
in {\sc Lepto} (dashed histograms) and {\sc Aroma} (dotted histograms).}
{fig:f2d3}

The resulting $x_\pom$-dependence from the model may be slightly  steeper than
in the data.  Fitting a universal $x_\pom^{-a}$-dependence we get $a=1.55$ 
from our model to be compared with the H1 result $a=1.2\pm 0.1$ \cite{H1-gaps}
and ZEUS results $a=1.3\pm 0.1$ \cite{ZEUS-gaps} and  $a=1.5\pm 0.1$
\cite{ZEUS-gaps2} using different methods. However the result obtained in the
model depends significantly on which  $x_\pom$-bins that are included in the
fitting procedure.  This originates from the fact that the $x_\pom$ dependence
in the extracted structure function does not give quite a straight line in this
plot,  but instead has a  tendency for a curvature with smaller slope at small
$x_\pom$ and  larger slope at large $x_\pom$. However, this is not a feature of
the model itself but rather a consequence of the selection criteria. Large
$x_\pom$ are suppressed due to the gap selection. Changing the  limit
$\eta_{\max}<3.2$ to $\eta_{\max}<4.2$  the slope changes from $a=1.55$ to
$a=1.25$.  The latter result is also similar to what is obtained if diffractive
events are selected by requiring a leading proton  (with $x_L>0.95$) which
gives a slope $a=1.24$. The resulting diffractive structure function with this
selection is also indicated with straight lines in Fig.~\ref{fig:f2d3}. We have
also applied the $M_X$ method \cite{ZEUS-gaps2} to select  diffractive events
which gives a slope $a=1.35$. This illustrates that there are large systematic
uncertainties  when extracting the $x_\pom$ dependence.

New data covering a larger $(\beta,Q^2)$ range have recently been showed at
conferences \cite{H1-eilat}, but the precise numbers are not yet available for
direct comparison with our model. The new data seem to indicate that the
$x_\pom$ dependence varies with $(\beta,Q^2)$. In our model the 
 $x_\pom$ dependence  is approximately the same for all $(\beta,Q^2)$
points. The underlying reason for this is that the  $x_\pom$ dependence in the
model is basically given by the  $\xi$ dependence of the gluon density 
($F_2^D(x_\pom) \propto x_\pom^{-1-d} \leftrightarrow \xi g(\xi)  \propto
\xi^{-d}$) as has also been noted in \cite{BH}. 

An important question is if the selection mechanism together with the
acceptance correction can influence the diffractive structure function. The
gap selection is dependent on Monte Carlo simulation to extract the efficiency
for detecting particles in the very forward region that would destroy the gap.
If this efficiency is overestimated then the result would be a too large
cross-section. Our model gives more activity in the very forward region than a
conventional pomeron based model and thus more gaps are destroyed by particles
from the remnant system. As an example of this, the diffractive structure
function obtained with our model increases with $\sim 30$ \% if the forward gap
limit is  decreased from $\eta=6.6$ to $\eta=5.6$.

There has been some speculation about the charm contribution  to the
diffractive structure function \cite{Stirling}. As Fig.~\ref{fig:f2d3} shows we
get a charm contribution of only a few percent in our model. We have also
implemented the SCI mechanism in the {\sc Aroma} Monte Carlo~\cite{AROMA},
which uses the BGF matrix element with explicit charm quark mass
included~\cite{GERHARD-heavyME}. This gives a factor $\sim 2$ lower charm rate
than in {\sc Lepto}, where the massless matrix element is used together
with a simple threshold factor. However, this difference is not due to
the different treatments of the matrix element but rather because in {\sc
Lepto} the charm sea is also included and gives an additional contribution.

\subsection{Transverse energy flows}

Our model is meant to be applicable for DIS in general and should
therefore also be compared with normal DIS events. The observed large
forward transverse energy flow in an inclusive event sample requires a
substantial energy and particle production and is thereby `orthogonal'
to forward rapidity gaps. In Fig.~\ref{fig:hadron_eflow9} we compare our model
result with the H1 data \cite{H1-eflow}. The agreement is quite good
except for the smallest $x$-values where the model is below the data in
the central region of the hadronic cms.

The soft colour interactions generates not only gap events, but also 
larger fluctuations in general. In particular, configurations may arise
where the string goes `back and forth' and thereby produces more energy per 
unit rapidity. The improved treatment of sea quark interactions
is also of relevance here, since it gives rise to a string from a valence
spectator parton to the partner sea quark remaining in the target system. 
Depending on the momentum of this sea quark partner, the corresponding 
string will extend more or less into the central region in rapidity, cf.
Figs.~\ref{fig:tflow} and \ref{fig:shower_ps}d. The 
hadronization of this string will thereby give another contribution to the 
forward energy flow. The effects of both these non-perturbative parts of 
the model are illustrated in Fig.~\ref{fig:hadron_eflow9} and it is clear that
they both give important contributions to an overall better description
of the data.  

\ffigw{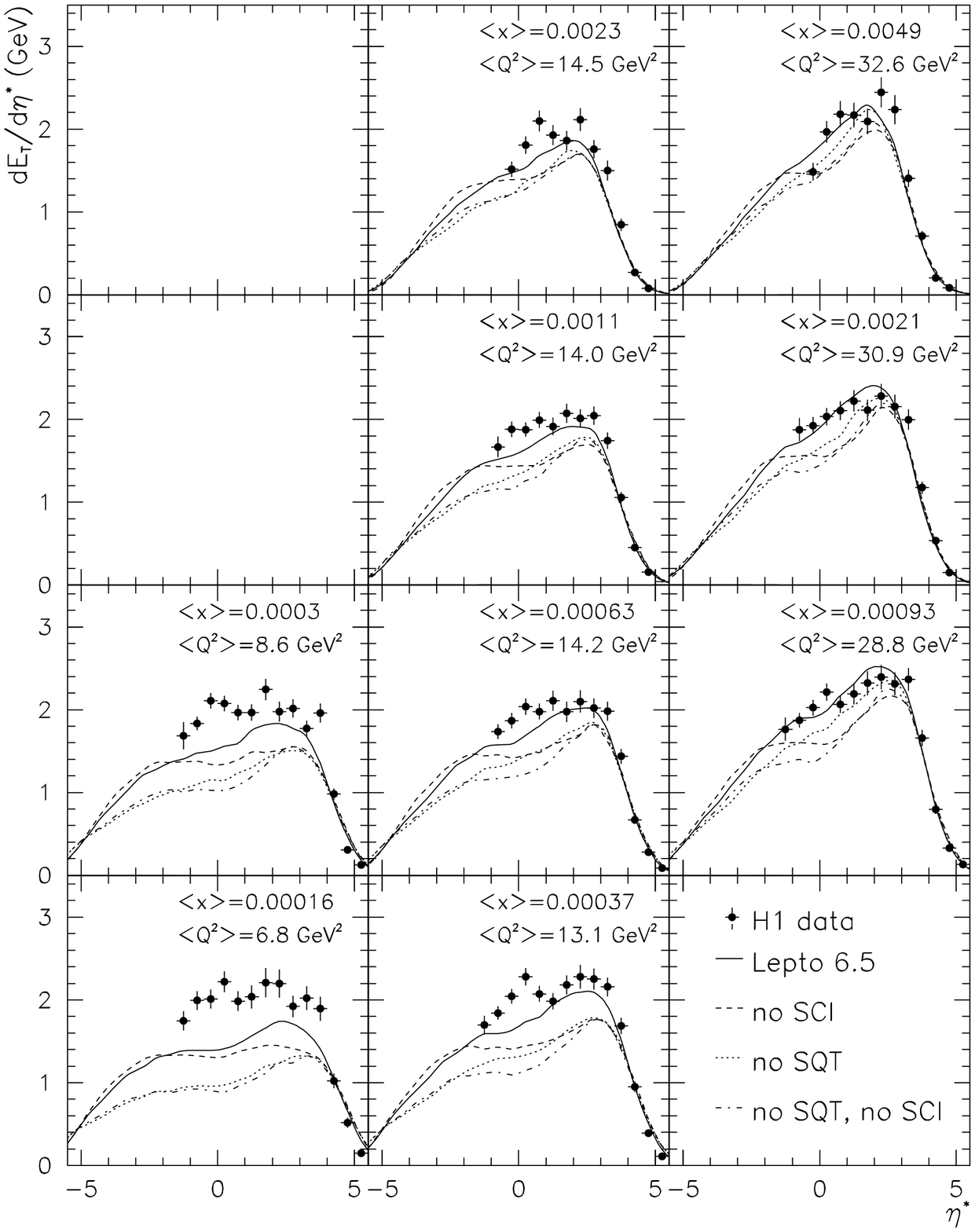}{14cm} {Transverse
energy flow versus pseudo-rapidity $\eta^*$ in the hadronic cms. H1 data points
\protect\cite{H1-eflow} compared to model curves:  the standard {\sc Lepto} 6.5
model (solid),  without soft colour interactions (dashed), without the new sea
quark treatment (dotted) and without both of these (dash-dotted).}
{fig:hadron_eflow9}

This also shows the importance of non-perturbative effects for the 
forward energy flow and makes it less suitable for detailed tests of the 
underlying perturbative dynamics in terms of, e.g., BFKL or GLAP. 
This was already demonstrated in Fig.~\ref{fig:tflow} above, in terms of the 
large difference between the parton and hadron level result. 

Finally, we note that the energy flows from the model are not sensitive to the
matrix element cut-offs in $z_q$ and $\hat{s}$, which can be varied
substantially  without giving a significant effect (not explicitly shown in
Fig.~\ref{fig:hadron_eflow9}). This is very satisfactory, since these cut-offs
are just a boundary between two different calculational  approximations, i.e.
the first order matrix elements and the parton showers. 

\section{Concluding discussion}
\label{sec:conclusions}

The observed large forward transverse energy flow has been argued  \cite{Golec}
to be a sign of BFKL dynamics as predicted in perturbative QCD at small-$x$. We
have here demonstrated that this observable is largely  influenced by
non-perturbative effects. In addition to normal hadronization,  there are also
significant effects related to the treatment of the proton  target remnant when
a sea quark interacted, which is more likely at small $x$,  as well as the soft
colour interactions. This makes is difficult to conclude  whether the observed
large forward transverse energy flow  should be taken as evidence for BFKL
behaviour. 

In our model we are using the conventional GLAP evolution
scheme \cite{GLAP} summing leading $\log{Q^2}$ terms. Terms with
$\log(1/x)$ are neglected in GLAP, but will become important at small
enough $x$ and should then be resummed as in the BFKL equation \cite{BFKL}.
Therefore GLAP evolution should no longer be valid at some small $x$.
Recent studies \cite{LDCM} based on the CCFM equation \cite{CCFM} 
which sums both leading $\log{Q^2}$ and $\log(1/x)$ terms
suggests that this happens only at very small $x$, below the region 
accessible at HERA. Thus, there is no strong reason that the GLAP formalism
should not be applicable for our purposes. Indeed, with the inclusion of  
the non-perturbative effects, our model is also able to reproduce the 
observed forward energy flow reasonably well.

Concerning the interpretation of the observed rapidity gap events, a 
major issue is whether only pomeron-based models are capable of describing
the data or other models can do it as well. 
Our model \cite{SCI} which is elaborated upon in this paper, 
is based on conventional DIS with perturbative QCD matrix elements and 
parton showers, but with the additional assumption of soft colour interactions 
before the normal Lund hadronization model is invoked. 

It should here be noted that other types of soft interactions have been
discussed in other contexts. Colour reconnections causing modified Lund string
topology has been investigated~\cite{StringReconn} in case of $e^+e^-
\rightarrow W^+W^- \rightarrow q_1\bar{q}_2q_3\bar{q}_4$, where the two
resulting strings may interact. Soft interactions of a colour charge moving
through a colour medium has been considered in~\cite{ColourMedium}, and argued
to give rise to large K-factors in Drell-Yan processes and synchrotron
radiation from the QCD vacuum. Although these studies are not related to
rapidity gaps, they contribute to a more general attempt to understand
non-perturbative QCD.

Returning to the rapidity gap problem, one may worry \cite{Leif-gaps} that
unsuppressed parton emission will populate the  rapidity region such as to
destroy the gaps, or lower their probability below  the observed one. The
cut-off in parton virtuality, defining the borderline to the non-perturbative
region, is a regulator of this. As discussed in section 2, our model has the
standard values for such cut-offs and there is no need for an extra suppression
of the perturbative emission. Nevertheless, we obtain  a significant rate of
gap events, in basic agreement with data. 

Of importance for the gap rate is also the fluctuations in the initial 
parton emission \cite{SCI}. Although, one may expect that a GLAP parton
shower gives a fair mean description of events, there is no guarantee
that it accounts properly for fluctuations. Larger fluctuations of the
number of emitted gluons would increase the rate of gap events  and
also  increase the inclusive forward energy flow due to `downwards'
and  `upwards' fluctuations, respectively.

If this new attempt to understand the rapidity gap phenomenon is successful 
in further detailed comparison with data, it will circumvent some
problems in the pomeron-based approach. In particular those associated
with the concept of a preformed exchanged object. This object cannot be
a real particle or state, since it has a negative mass square $t$. It
could be a virtual exchange corresponding to some real state, such as a
glueball, but this is presently unclear (although a recent glueball
candidate \cite{WA91} fits on the pomeron trajectory). The
interpretation of factorization in Regge phenomenology in terms of an
emission of a pomeron given by a pomeron flux and a pomeron-particle 
interaction cross-section also has some problems. Since it is only
their product that is experimentally observable one cannot, without
further assumptions, define the absolute normalization of this flux and
cross-section unambiguously \cite{Landshoff}. This also gives a
normalization ambiguity for the parton densities of the pomeron which
is reflected in the problem of whether they fulfill a momentum sum rule
or not. 

Leaving the concept of a preformed object and instead considering
\cite{Soper} a process with interactions with the proton both before
and after the hard scattering (taking place at a short space-time
scale) may avoid these problems, as in our model. One may ask whether
the soft colour interactions introduced here is essentially a model for
the pomeron. This should not be the case as long as no pomeron or Regge
dynamics is introduced in the model. 
At present, the model has only one new free parameter, the colour exchange
probability $R$. Other parameters belong to the conventional DIS model
\cite{LEPTO65} and have their usual values. 

Our model attempts to describe both diffractive and inclusive interactions
in a unified framework. Thus, data on both aspects should be used to
test it. Typical diffractive characteristics concern
properties of the forward $R$-system and the $M_X$-distribution. 
Here, one should investigate the dependence on the size of the
rapidity gap and leading particle requirements. Pomeron-based models,
which are constructed with a leading proton and a gap, are likely to be
less sensitive compared to our model, where the gap size depends on
fluctuations in parton emissions and soft interactions and the
$R$-system is less constrained.

In conclusion, we have shown that it is possible to describe two very 
different observations, namely large forward energy flows in DIS and a 
substantial fraction of events with a large rapidity gap, within one unified 
framework based on standard perturbative QCD and hadronization plus the novel
assumption of soft colour interactions.

\end{document}